\documentclass[onecolumn,showpacs,pra,longbibliography]{revtex4-1}
\usepackage{graphicx}
\usepackage{latexsym}
\begin{document}

\title{Effect of noisy channels on the transmission of mesoscopic twin-beam states}
% Coherence waves in evolving intense twin beams

\author{Alessia Allevi}
\affiliation{Department of Science and High Technology, University of Insubria and Institute for Photonics and Nanotechnologies, IFN-CNR, Via Valleggio 11, I-22100 Como, Italy}

\author{Maria Bondani}
\affiliation{Institute for Photonics and Nanotechnologies, IFN-CNR, Via Valleggio 11, I-22100 Como, Italy}

\begin{abstract}
Quantum properties of light, which are crucial resources for Quantum Technologies, are quite fragile in nature and can be degraded and even concealed by the environment. 
We show, both theoretically and experimentally, that mesoscopic twin-beam states of light can preserve their nonclassicality even in the presence of major losses and different types of noise, thus suggesting their potential usefulness to encode information in Quantum Communication protocols.
We develop a comprehensive general analytical model for a measurable nonclassicality criterion and find thresholds on noise and losses for the survival of entanglement in the twin beam. 
\end{abstract}

\maketitle

%%%%%%%%%%%%%%%%%%%%%%%%%%  body  %%%%%%%%%%%%%%%%%%%%%%%%%%
\section{Introduction}
The proper transmission of quantum states of light is at the core of Quantum Communication protocols.
Among the states involved in such protocols, there are bipartite states exhibiting nonclassical correlations, such as entangled states. Communication protocols can be performed through optical fibers \cite{Mul95, Mar03, Urs04}, free-space channels \cite{Kur02, Asp03, Jin10}, and underwater environment \cite{Ji17, Huf20}.
As extensively theoretically discussed and experimentally proved \cite{Vasy12,Capraro12}, quantum states of light are rather fragile since their degree of nonclassicality can be deteriorated or even concealed by the losses that affect the transmission channels. For instance, the transmission through free-space channels is limited by the presence of atmospheric turbulence, which acts as a temporal and spatial variation of the air refraction index \cite{Boh17}. Usually Quantum Communication experiments have been implemented by exploiting different degrees of freedom, such as polarization, time-bin, position and transverse momentum \cite{Cozzolino19, Flamini19}, all of them at the single-photon level  \cite{But98, Mir15, Jin19}.\\
In some recent papers of ours \cite{JOSAB19,ApplSci20} we have demonstrated that quantum optical states produced in the mesoscopic intensity regime are potentially useful to accomplish Quantum Communication protocols as they exhibit a good robustness to different  statistically-distributed losses. 
In principle, a similar situation occurs also in the presence of noise sources, which characterize the transmission channel itself or can be added by some evesdropper.
Indeed, a noise source, thermal or multi-mode thermal one, can add an uncorrelated amount of noise to the transmitted quantum optical state thus resulting detrimental for its nonclassical nature \cite{Weed10}.
In this respect, we demonstrate here that the more superPoissonian the noise, the more detrimental its effect on the transmitted quantum state. On the contrary, the more subPoissonian the noise, the less detrimental its effect on the transmitted quantum state.  
In the following, we focus on the case of a noise source added to one of the parties of a mesoscopic twin-beam (TWB) state and theoretically demonstrate the way in which it modifies the nonclassicality of the transmitted quantum state. Depending on the nature of the noise source, the quantum correlations exhibited by the TWB can slightly decrease or be completely destroyed. In particular, we consider some limiting situations that can be experimentally addressed.
We show experimental results obtained by mixing two different noise sources, a Poissonian or a single-mode thermal state, with a multi-mode TWB and evaluate the level of nonclassicality of the resulting state in terms of the noise reduction factor. By applying the developed theoretical model, we also show that we are able to characterize either the noise or the quantum state.     
\section{Modelling the noise affecting the transmission channel}
\begin{figure} % Fig. 1
 \centering
\includegraphics[width=\linewidth]{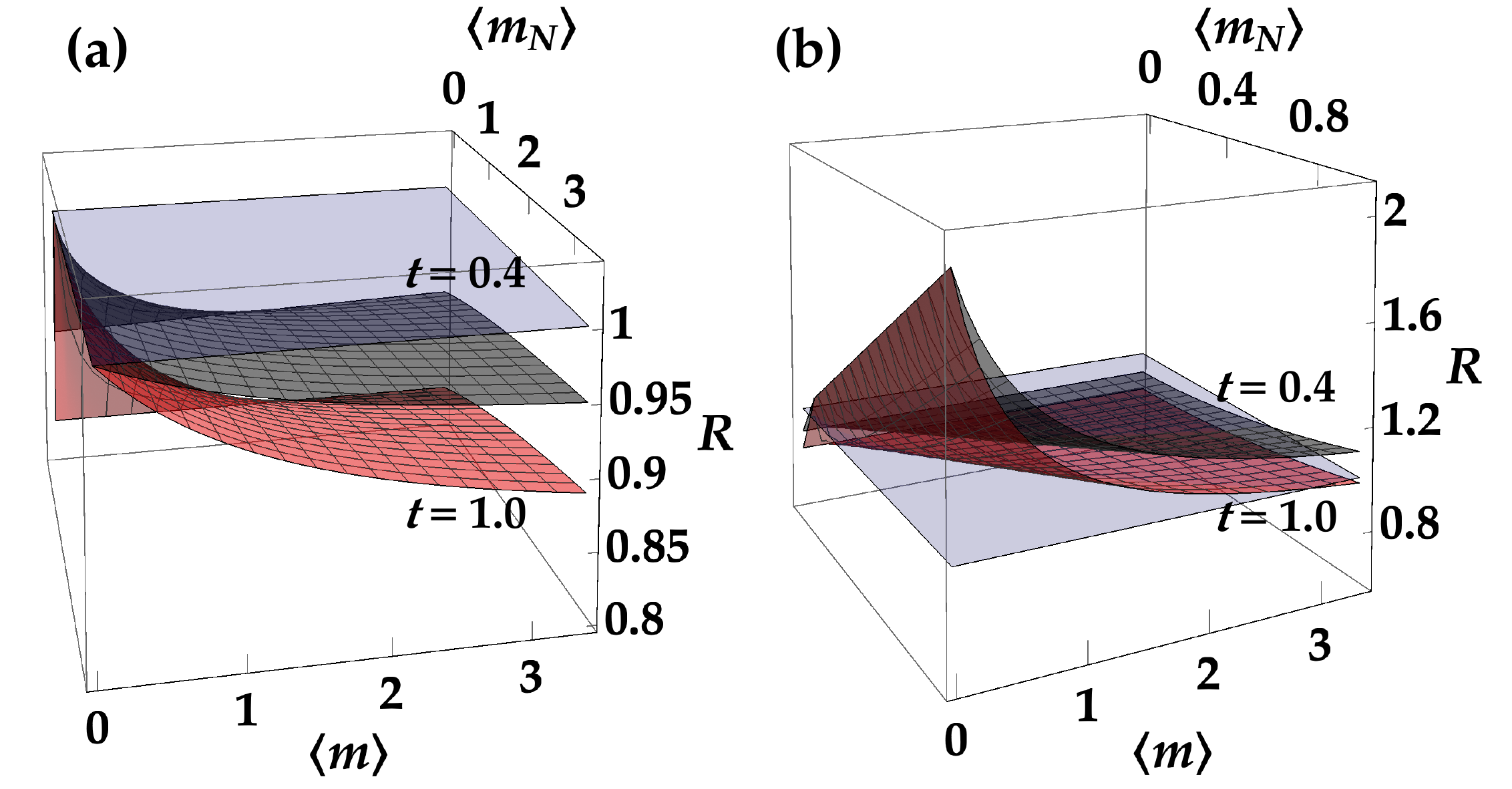}
 \caption{Noise reduction factor as a function of the mean value of TWB and of the mean value of noise when it is described by a Poissonian distribution (panel (a)) or by a single-mode thermal state (panel (b)). In both cases we set $\mu =100$, $\eta =0.17$ and two possible values of the transmittance efficiency: $t = 1$ (red surface) and $t = 0.4$ (black surface). The blue surface at $R = 1$ represents the boundary plane between classical and nonclassical correlations.}
\label{figPoissTherm}
\end{figure}

\begin{figure} % Fig. 1
 \centering
\includegraphics[width=\linewidth]{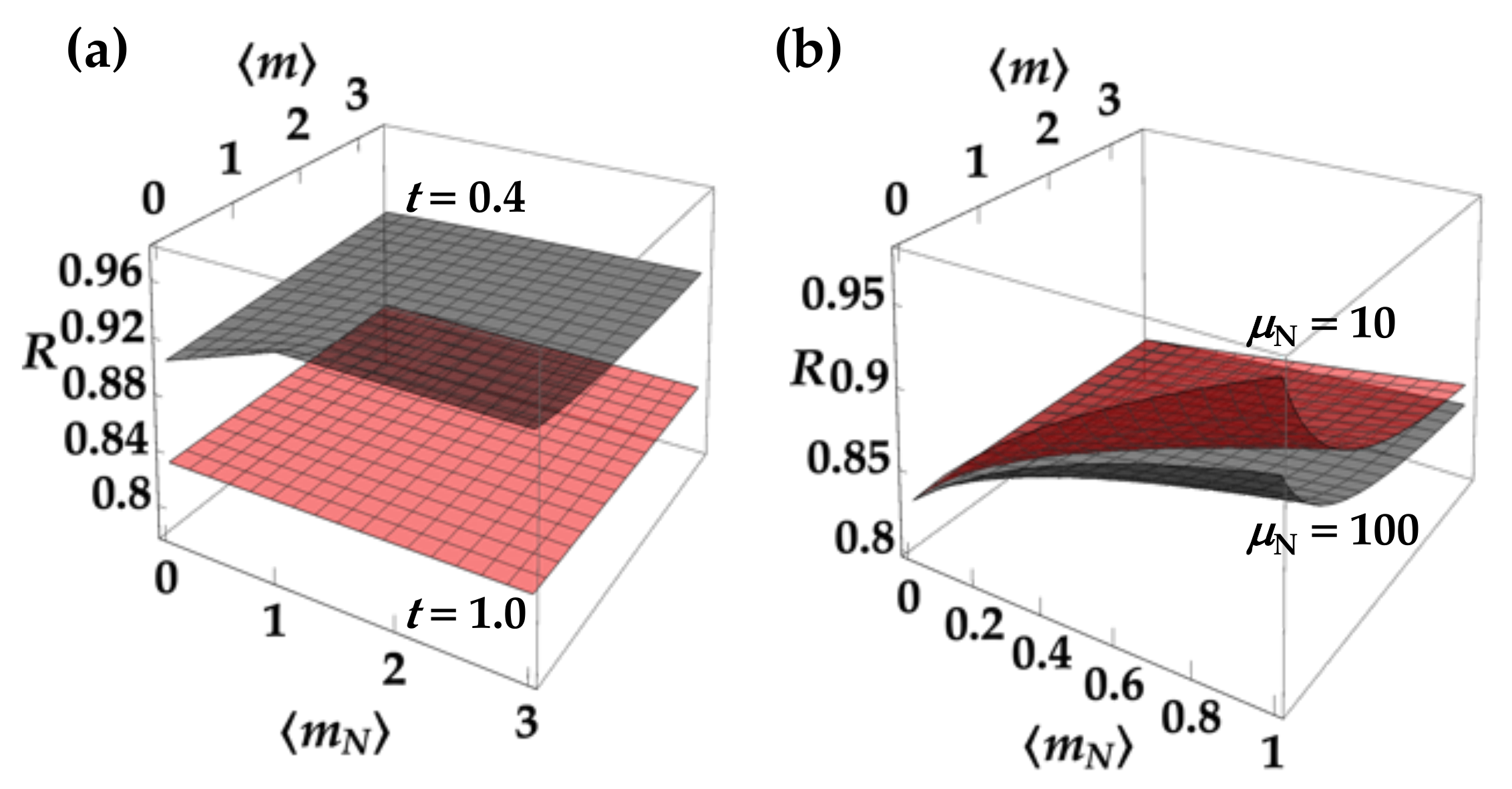}
 \caption{Noise reduction factor as a function of the mean value of TWB and of the mean value of noise when it is described by a Fock distribution with $t = 1$ (dark gray surface) and $t= 0.4$ (red surface) in panel (a) and by a conditional state with  $t= 1$, $m_{\rm cond} = 5$ and $\mu_{\rm N} = 100$ (dark gray surface) or $\mu_{\rm N} = 10$ (red surface) in panel (b). In both cases we set $\mu = 100$, $\eta = 0.17$.}
\label{figFockCond}
\end{figure}

Let us assume that the bipartite state to be transmitted is a multi-mode TWB state described as the tensor product of $\mu$ identical ($i.e.$ equally populated) TWB states \cite{arimondo}:
\begin{equation} \label{multiTWB}
| \psi_{\mu} \rangle = \sum_{n=0}^{\infty} \sqrt{P^{\mu}(n)}|n^{\otimes}\rangle |n^{\otimes}\rangle,
\end{equation}
where $|n^{\otimes}\rangle =\delta(n- \sum_{h=1}^{\mu}n_h) \bigotimes_{k=1}^{\mu} |n\rangle_k$ is the overall number of photons
in the $\mu$ modes that impinge on the detector, while $P^{\mu}(n)$ is the multi-mode thermal distribution 
\begin{equation}
P^{\mu}(n) = \frac{(n+ \mu -1)!}{n! (\mu -1)!\left(\langle n \rangle+1\right)^{\mu}\left(1/\langle n \rangle+1\right)^n},
\end{equation}
in which $\langle n \rangle$ is the mean number of photons per mode in each arm.
The state in Eq.~(\ref{multiTWB}) is an entangled state in the number of photons. To experimentally prove its nonclassical nature, a sufficient criterion for entanglement is given by the noise reduction factor \cite{pra07}, defined as  
\begin{equation} \label{Rphoton}
R = \frac{\sigma^2(n_1-n_2)}{\langle n_1 \rangle + \langle n_2 \rangle},
\end{equation}
where $\sigma^{2}(n_1 - n_2)$ is the variance of the distribution of the photon-number difference between the two parties, while $\langle n_1 \rangle + \langle n_2 \rangle$ is the shot-noise-level, that is the value of $\sigma^2(n_1 - n_2)$ in the case of two coherent states having mean values $\langle n_1 \rangle$ and $\langle n_2 \rangle$.
We have already demonstrated that the expression of $R$ in Eq.~(\ref{Rphoton}), which is given in terms of photons, can be also written in terms of measurable quantities to be directly applied to data \cite{jointdiff}.
In particular, in the case of multi-mode thermal TWBs, Eq.~(\ref{Rphoton}) reads \cite{EPJD18}
\begin{equation} \label{Rmulti}
R = 1- \frac{2\sqrt{\eta_1 \eta_2}\sqrt{\langle m_1 \rangle \langle m_2 \rangle}}{\langle m_1 \rangle + \langle m_2 \rangle} + \frac{(\langle m_1 \rangle- \langle m_2 \rangle)^2}{\mu (\langle m_1 \rangle+ \langle m_2 \rangle)}.
\end{equation}
Now, if we assume that both the arms of TWB are affected by an uncorrelated amount of noise having mean value $\langle m_{\rm N_1}\rangle$ and variance $\sigma^2(m_{\rm N_1})$ on arm 1 and $\langle m_{\rm N_2}\rangle$ and variance $\sigma^2(m_{\rm N_2})$ on arm 2,  the expression in Eq.~(\ref{Rmulti}) must be re-calculated. We consider the single-shot values  $k_{1,2} = m_{1,2} + m_{\rm N_{1,2}}$ and evaluate the noise reduction factor $R=\sigma^{2}(k_1 - k_2)/\langle k_1 + k_2 \rangle$. We get
\begin{eqnarray} \label{noisyRdouble}
R &=& 1- \frac{2\sqrt{\eta_1 \eta_2} \sqrt{\langle m_1 \rangle \langle m_2 \rangle}}{\langle m_1 \rangle + \langle m_2 \rangle + \langle m_{\rm N_1} \rangle+ \langle m_{\rm N_2} \rangle} \\
&+& \frac{(\langle m_1 \rangle- \langle m_2 \rangle)^2}{\mu (\langle m_1 \rangle+ \langle m_2 \rangle + \langle m_{\rm N_1} \rangle+ \langle m_{\rm N_2} \rangle)} + \frac{\sigma^2(m_{\rm N_1})-\langle m_{\rm N_1} \rangle+\sigma^2(m_{\rm N_2})-\langle m_{\rm N_2} \rangle}{\langle m_1 \rangle+ \langle m_2 \rangle + \langle m_{\rm N_1} \rangle+ \langle m_{\rm N_2} \rangle}. \nonumber
\end{eqnarray}
From Eq.~(\ref{noisyR}) it emerges that the additional noise comes into play in the last term as well as in each denominator. 
An interesting aspect of Eq.~(\ref{noisyR}) is that the last term can be positive, negative or zero depending on the nature of the noise. 
To study the feasible situation in which one arm of TWB is sent through a transmission channel to the receiver, while the other arm is maintained by the sender, in the following we focus on the case where only in one arm of TWB, say arm 2, there is an additional noise source ($\langle m_{\rm N_2} \rangle \equiv \langle m_{\rm N} \rangle$), so that Eq.~(\ref{noisyRdouble}) simplifies to
\begin{equation} \label{noisyR}
R = 1- \frac{2\sqrt{\eta_1 \eta_2} \sqrt{\langle m_1 \rangle \langle m_2 \rangle}}{\langle m_1 \rangle + \langle m_2 \rangle + \langle m_{\rm N} \rangle} 
+ \frac{(\langle m_1 \rangle- \langle m_2 \rangle)^2}{\mu (\langle m_1 \rangle+ \langle m_2 \rangle + \langle m_{\rm N} \rangle)} + \frac{\sigma^2(m_{\rm N})-\langle m_{\rm N} \rangle}{\langle m_1 \rangle+ \langle m_2 \rangle + \langle m_{\rm N} \rangle}. 
\end{equation}
If the transmission channel is also affected by an asymmetric amount of loss such that $\langle m_1 \rangle = \langle m \rangle = \eta \langle n \rangle$, $\langle m_2 \rangle = t \langle m \rangle = t \eta \langle n \rangle$, $t \in (0,1)$ being the transmission efficiency, Eq.~(\ref{noisyR}) becomes
\begin{equation} \label{noisylossyR}
R = 1- \frac{2\eta t \langle m \rangle}{(1+t) \langle m \rangle + \langle m_{\rm N} \rangle} 
+ \frac{(1-t)^2 \langle m \rangle^2}{\mu \left[(1+t) \langle m \rangle + \langle m_{\rm N} \rangle \right]} + \frac{\sigma^2(m_{\rm N})-\langle m_{\rm N} \rangle}{(1+t)\langle m \rangle + \langle m_{\rm N} \rangle}. 
\end{equation}
It is easy to show that in the case of a coherent noise contribution for which $\sigma^2(m_{\rm N}) = \langle m_{\rm N} \rangle$, Eq.~(\ref{noisylossyR}) reduces to the case of a TWB affected by an asymmetric amount of noise \cite{ApplSci20} apart from the presence of $\langle m_{\rm N} \rangle$ in the denominators:
\begin{equation} \label{noisyRcoh}
R = 1- \frac{2\eta t \langle m \rangle}{(1+t) \langle m \rangle + \langle m_{\rm N} \rangle} + \frac{(1-t)^2 \langle m \rangle^2}{\mu \left[(1+t) \langle m \rangle + \langle m_{\rm N} \rangle \right]}. 
\end{equation}
In such a case, the typical behavior of $R$ as a function of the mean value of TWB and of the mean value of the Poissonian noise source is shown in panel (a) of Fig.~\ref{figPoissTherm} for $t = 1$ (red surface) and $t = 0.4$ (black surface).\\
On the contrary, if the noise contribution is multi-mode thermal with $\mu_{\rm N}$ equally-populated modes for which $\sigma^2(m_{\rm N}) = \langle m_{\rm N} \rangle (\langle m_{\rm N} \rangle/\mu_{\rm N} +1) > \langle m_{\rm N} \rangle$ \cite{ASL}, Eq.~(\ref{noisylossyR}) becomes  
\begin{equation} \label{noisyRth}
R = 1- \frac{2\eta t \langle m \rangle}{(1+t) \langle m \rangle + \langle m_{\rm N} \rangle} 
+ \frac{(1-t)^2 \langle m \rangle^2}{\mu \left[(1+t) \langle m \rangle + \langle m_{\rm N} \rangle \right]} + \frac{\langle m_{\rm N} \rangle ^2}{\mu_{\rm N}\left[(1+t)\langle m \rangle + \langle m_{\rm N} \rangle\right]}. 
\end{equation}
Note that the last term becomes negligible for large numbers of modes, while it can play a detrimental role for small values of $\mu_{\rm N}$, being the worst for $\mu_{\rm N} = 1$. The expected behavior for a noise contribution described by a single-mode thermal state ($\mu_{\rm N} = 1$) is shown in panel (b) of Fig.~\ref{figPoissTherm} for $t = 1$ (red surface) and $t = 0.4$ (black surface).
It is relevant to explore the presence of a multi-mode thermal noise source since it is expected in standard communication channels \cite{Weed10}. \\
However, for the sake of completeness, we will also explore the case of a subPoissonian noise, which could be intentionally introduced by some evesdropper. 
The variance of a detected Fock state under Bernoullian detection is $\sigma^2 (m_{\rm N}) = \eta^2 \sigma^2(n) + \eta (1-\eta) \langle n \rangle$ \cite{jointdiff} with $\sigma^2(n) = 0$. Thus the expression in Eq.~(\ref{noisylossyR}) reads as
\begin{equation} \label{noisyRfock}
R = 1- \frac{2\eta t \langle m \rangle}{(1+t) \langle m \rangle + \langle m_{\rm N} \rangle}
+ \frac{(1-t)^2 \langle m \rangle^2}{\mu \left[(1+t) \langle m \rangle + \langle m_{\rm N} \rangle \right]} - \frac{\eta t \langle m_{\rm N} \rangle}{(1+t)\langle m \rangle + \langle m_{\rm N} \rangle}, 
\end{equation}
where the noise contribution is in arm 2.
In this case it is straightforward to prove that, for $t = 1$, the Fock state does not change the value of $R$, which, at variance, increases for $t<1$. In particular, for multi-mode TWB states with a large number of modes ($\mu \sim 100$), the value of $R$ saturates to $(1-\eta t$), as shown in panel (a) of Fig.~\ref{figFockCond}.\\
Another example of subPoissonian noise contribution is given by the conditional states obtained in one arm of TWB (say the signal) when conditional measurements are performed in the other arm (the idler) \cite{EPL,lau03,our06,perina13,isk16}. In such a case, the mean value and the variance of the noise contribution can be written as functions of the parameters describing the original state \cite{lamperti,OL19}:
\begin{eqnarray}
\langle m_{\rm N} \rangle &=& \frac{m_{\rm cond} (\langle m_{\rm TWB,N} \rangle +\eta_2 \mu_{\rm N})+\mu_{\rm N} \langle m_{\rm TWB,N} \rangle (1-\eta_2)}{\langle m_{\rm TWB,N} \rangle+\mu_{\rm N}}\\
\sigma^2(m_{\rm N}) &=& \frac{(1-\eta_2)}{(\langle m_{\rm TWB,N} \rangle+\mu_{\rm N})^2}[\eta_2 m_{\rm cond} \mu^2_{\rm N} + \langle m_{\rm TWB, N} \rangle \mu_{\rm N} (m_{\rm cond}+2 \eta_2 m_{\rm cond}+\mu_{\rm N})\nonumber\\
&+& \langle m_{\rm TWB,N} \rangle^2 (2 m_{\rm cond}+2\mu_{\rm N}-\eta_2\mu_{\rm N})],
\end{eqnarray} 
$\langle m_{\rm TWB,N} \rangle$ and $\mu_{\rm N}$ being the mean value and the number of modes of the unconditioned state, respectively, and $m_{\rm cond}$ the conditioning value. Even if this kind of subPoissonian noise degrades the level of nonclassicality of the transmitted TWB state, its effect is definetely less detrimental than that of Poissonian or superPoissonian noise sources, as shown in panel (b) of Fig.~\ref{figFockCond} for $t = 1$ and two different choices of $\mu_{\rm N}$.
 \section{Extracting information on the noise affecting the transmission channel}
\begin{figure}  
\centering
\includegraphics[width=\linewidth]{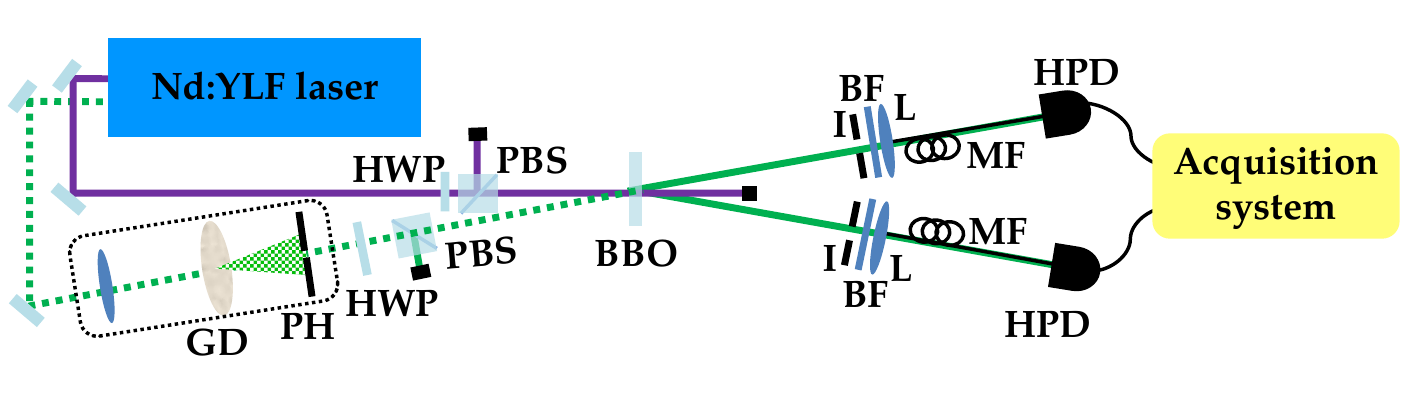}
 \caption{Sketch of the experimental setup. See the text for details.} 
\label{setup}
\end{figure}

\begin{figure} 
\centering
\includegraphics[width=11cm]{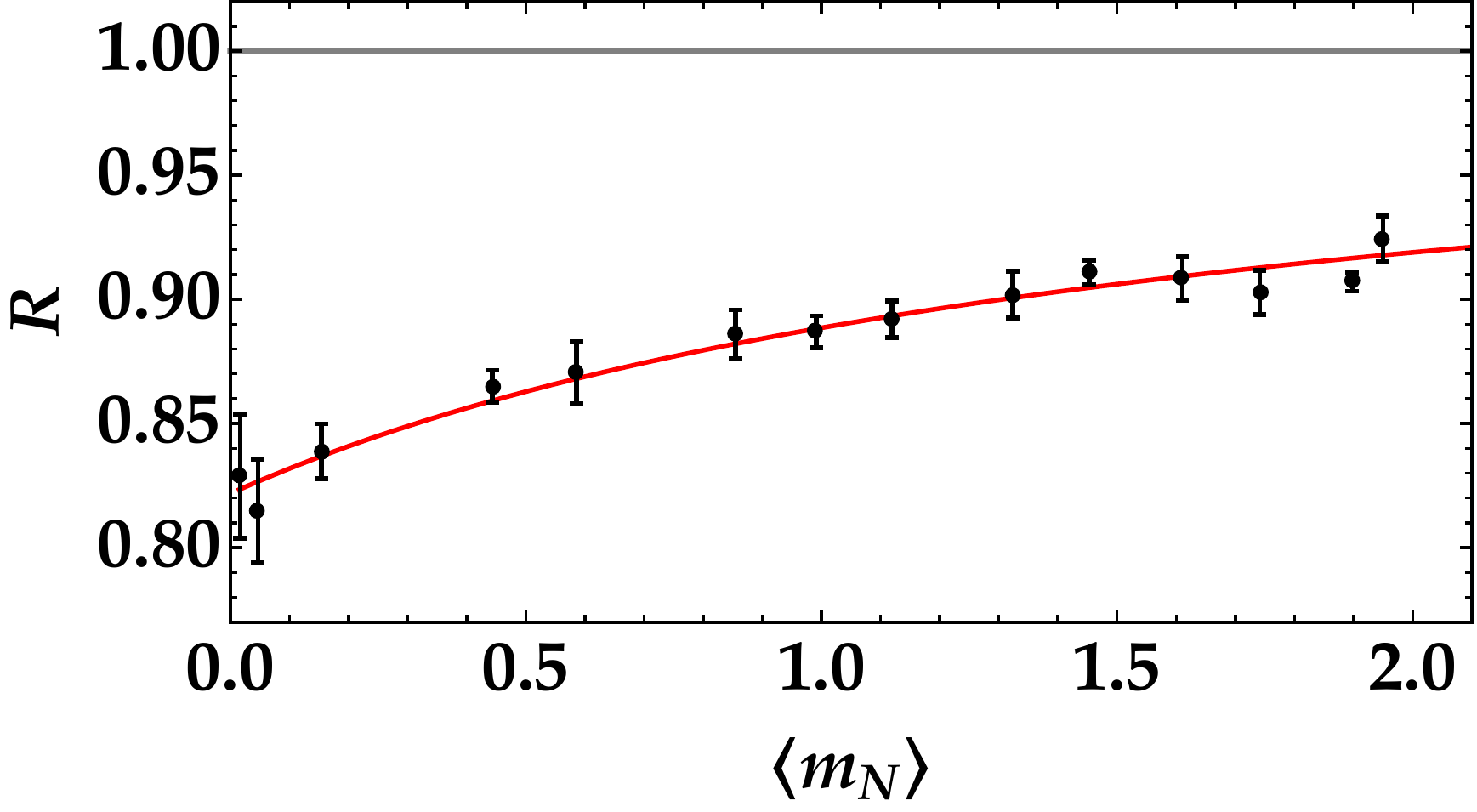}
 \caption{Noise reduction factor as a function of the mean number of detected photons of coherent noise for a fixed choice of TWB. Black dots and error bars: experimental data; red curve: best fit according to Eq.~(\ref{noisyRcoh}). Fitting parameters: $\langle m \rangle = 0.88$, $\mu = 1564$, $\eta = 0.19$, and $t = 0.90$; $\chi^{2}_{\nu}$ per degree of freedom: 1.12. The gray line at $R = 1$ represents the boundary between classical and nonclassical correlations.} 
\label{RvsPoiss}
\end{figure}

\begin{figure} 
 \centering
\includegraphics[width=11cm]{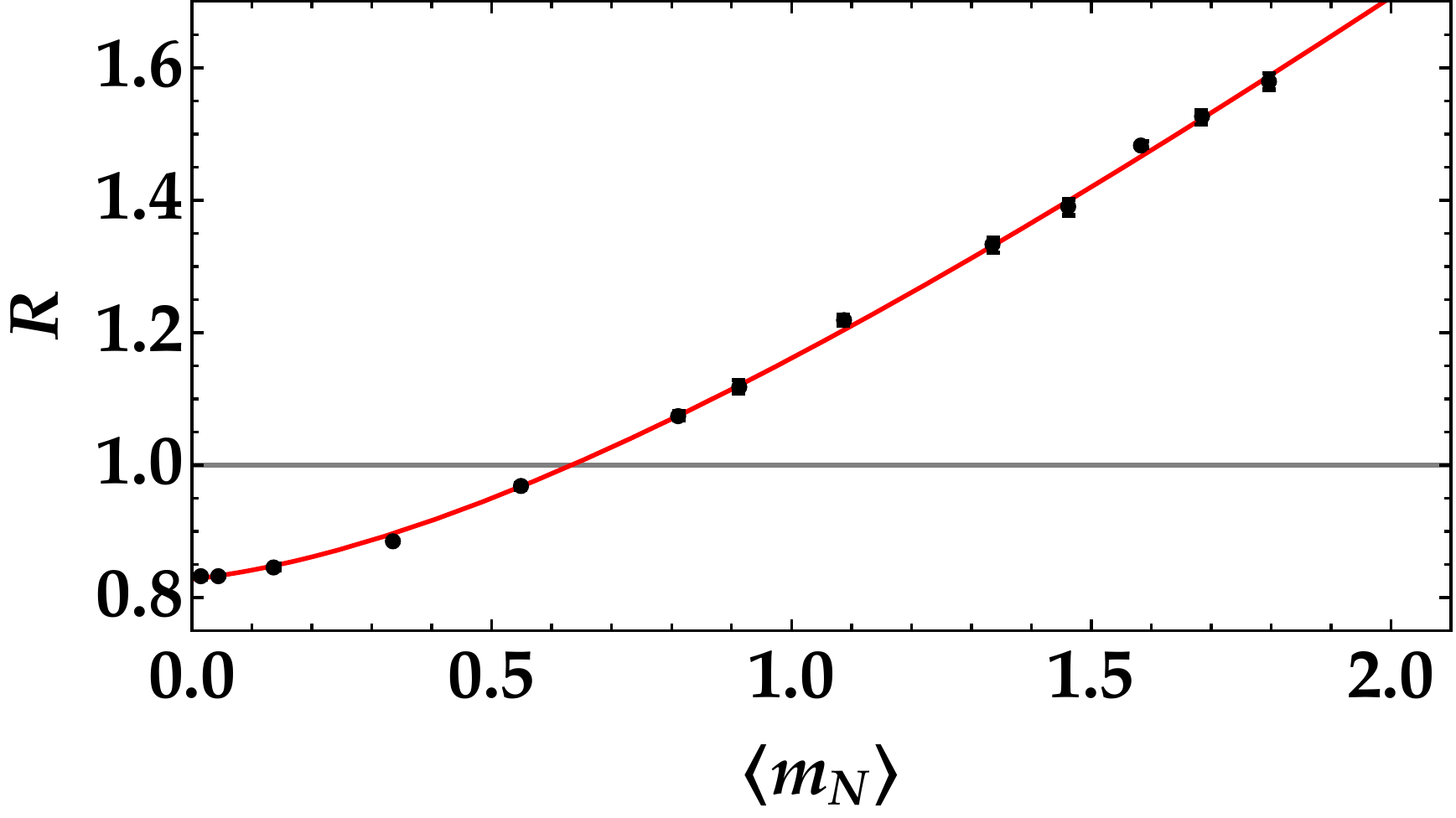}
 \caption{Noise reduction factor as a function of the mean number of detected photons of multi-mode thermal noise for a fixed choice of TWB. Black dots and error bars: experimental data; red curve: best fit according to Eq.~(\ref{noisyRth}). Fitting parameters: $\langle m \rangle = 0.89$, $\mu = 49.93$, $\eta = 0.18$, $t = 0.90$, and $\mu_{\rm N} = 1.38$; $\chi^{2}_{\nu}$ per degree of freedom: 2.21. The gray line at $R = 1$ represents the boundary between classical and nonclassical correlations.} 
\label{RvsTh}
\end{figure}

\begin{figure} 
 \centering
\includegraphics[width=11cm]{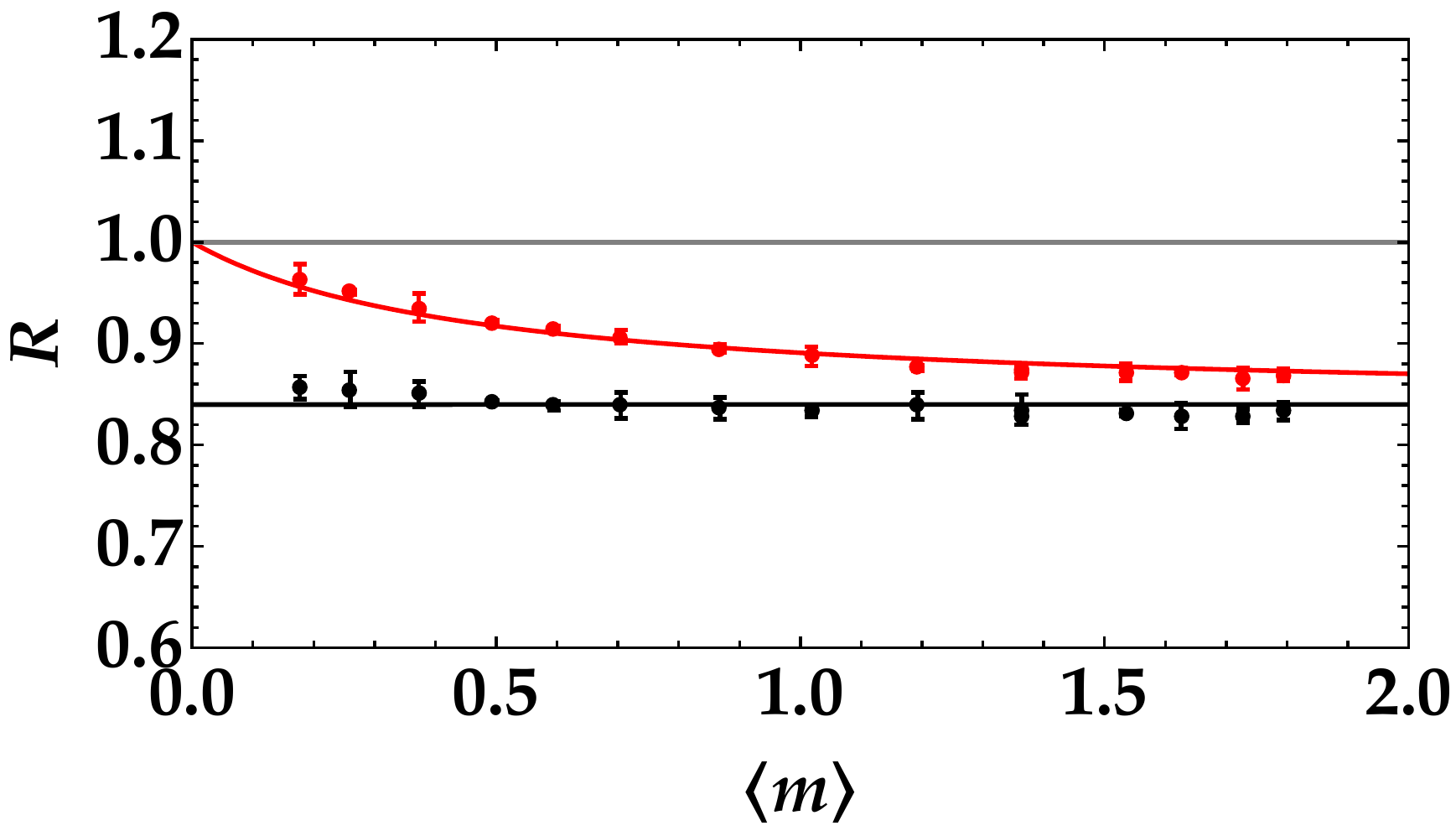}
 \caption{Noise reduction factor as a function of the mean number of detected photons of TWB state for a fixed choice of coherent noise. Black dots and error bars: experimental data in the absence of noise; red dots and error bars: experimental data in the presence of coherent noise; colored curves: theoretical fitting functions according to Eq.~(\ref{noisyRcoh}). From the fitting procedure applied to black dots we obtain: $\mu = 643.12$, $\eta = 0.17$, $t = 0.87$. By using the same values, from that applied to red dots we get $\langle m_{\rm N} \rangle = 0.87$.  The $\chi^{2}_{\nu}$ per degree of freedom is equal to 1.9 (black dots) and 3.9 (red dots). The gray line at $R = 1$ represents the boundary between classical and nonclassical correlations.} 
\label{RvsTWBPoiss}
\end{figure}

In the previous Section, we have demonstrated that, in general, the presence of an additional noise in one arm of the transmitted TWB state determines a degradation of its nonclassicality. In particular, we have proven that, the more superPoissonian the noise contribution, the worse the effect on TWB.
We prove this statement, by presenting the experimental results obtained for two different kinds of noise,  Poissonian and quasi-single-mode thermal, added to a component of a TWB. We show that the measurement of the noise reduction factor of the transmitted TWB allows us to extract information on the transmitted state as well as on the main properties of the noise source.\\
The setup used to produce the multi-mode TWB states of Eq.~(\ref{multiTWB}) is shown in Fig.~\ref{setup}.
The fourth-harmonic field at 262 nm of a Nd:YLF laser regeneratively amplified at 500 Hz (IC-500, High-Q Laser) is used to produce multi-mode TWB states in a slightly non-collinear interaction geometry by parametric down conversion in a $\beta$-barium-borate (BBO, cut angle = 46.7 deg, 6-mm long) crystal. The twin portions at frequency degeneracy (523 nm) are spatially and spectrally selected, focused by achromatic doublets into multi-mode fibers with a 600-$\mu$m-core diameter and thus delivered to the detectors.
In one arm of the TWB, a mirror is inserted in front of the BBO to introduce the noise contribution. As the first case, we use the second harmonics of the laser, which is a Poissonian light source at the same frequency as the TWB, while in the second case we use a quasi-single-mode thermal state produced by sending the second harmonics of the laser through a rotating ground-glass disk (GD) and selecting a single speckle with a pin-hole (PH).\\
Since we are interested in exploring the so-called mesoscopic intensity domain, where the optical states contain some numbers of photons, to prove their robustness to noise sources, the employed detectors are endowed with photon-number-resolving capability. In particular, they are hybrid photodetectors (HPD, mod. R10467U-40, Hamamatsu Photonics), a class of commercial detectors with partial photon-number resolution and a good linearity. Each detector output is amplified, synchronously integrated and digitized. 
By applying the self-consistent method already explained elsewhere \cite{ASL,arimondo}, we can convert the output voltages into numbers of detected photons and thus calculate all the relevant statistical quantities, such as the noise reduction factor, in terms of measurable quantities \cite{pra12}.  
In the following we present two different experiments: first of all, we show the behavior of the noise reduction factor as a function of the mean value of the noise contribution while keeping fixed the mean value of the transmitted TWB. Secondly, we do the reverse, that is we keep the mean number of photons of  the noise contribution fixed and we vary the mean number of photons of TWB.
We modify the energy of the TWB by means of a half-wave plate followed by a polarizing cube beam splitter on the pump beam, whereas we change the energy of the noise source by placing the same kinds of optical elements (half-wave plate and polarizing cube beam splitter) on its pathway (see Fig.~\ref{setup}). \\
%In particular, the half-wave plate on the pump beam (at 262 nm) is rotated in steps of 2.5 deg, while that on the TWB (at 523 nm) in steps of 2 deg.\\
In Fig.~\ref{RvsPoiss} we show the behavior of the noise reduction factor $R$ as a function of the mean number of photons of the Poissonian noise contribution for a fixed choice of the TWB state. The experimental data are shown as black dots + error bars, while the best fit according to Eq.~(\ref{noisyRcoh}) is shown as a red curve. The fit also takes into account the presence of an imbalance between the two arms of TWB. The fitting parameters are: the mean number of detected photons $\langle m \rangle$ and the number of modes $\mu$ of TWB, the quantum efficiency $\eta$, and the transmittance efficiency $t$. 
On observing the data, we notice that the presence of a Poissonian noise degrades the level of nonclassicality, even if the values remain well below 1.\\
The situation is completely different in the case of a thermal-noise source. Indeed, in Fig.~\ref{RvsTh} it is possible to appreciate the rapid growth of the noise reduction factor as a function of the mean number of detected photons of the added noise.
The data are presented as black dots + error bars, while the best fit is shown as a red curve. In this case, among the fitting parameters we have to include the number of modes of the noise contribution $\mu_{\rm N}$.   
In such a condition, it emerges that when the mean number of detected photons of the thermal contribution is less than that of the TWB it is still possible to observe nonclassicality, while for larger mean values the noise reduction factor goes well above 1. Further discussions about this point are addressed at the end of the Section.\\
To completely validate the model presented in Eq.~(\ref{noisyR}), in the following we show the results obtained by fixing the mean value of the noise and varying that of the TWB state.
In particular, in Fig.~\ref{RvsTWBPoiss} we present the noise reduction factor as a function of the mean value of TWB in the case of a coherent noise source in one arm. For the sake of comparison, in the same figure we plot the data in the absence (black dots + error bars) and in the presence (red dots + error bars) of the noise source. It is evident that the noise increases the value of $R$. However, it is interesting to notice that at increasing mean values of TWB, the nonclassicality degradation becomes more and more negligible.  
%%%%%%%%%%%%%%%%%%%%%%%%%%
\begin{figure} 
 \centering
\includegraphics[width=11cm]{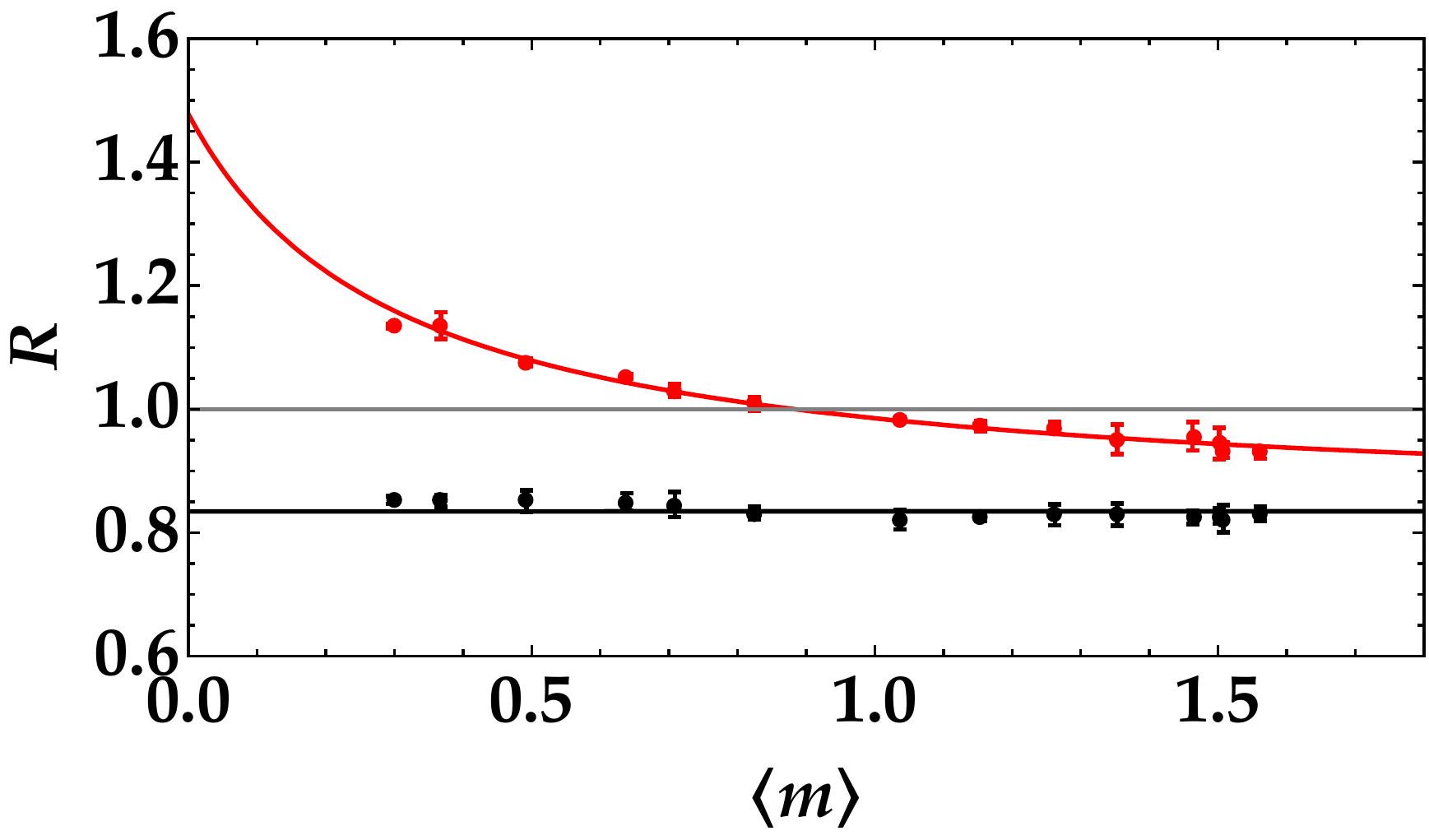}
 \caption{Noise reduction factor as a function of the mean number of detected photons of TWB state for a fixed choice of multi-thermal noise. Black dots and error bars: experimental data in the absence of noise; red dots and error bars: experimental data in the presence of thermal noise; colored curves: theoretical fitting functions according to Eq.~(\ref{noisyRth}). From the fitting procedure applied to black dots we obtain: $\mu = 499.48$, $\eta = 0.18$, $t = 0.87$. By using the same values, from that applied to red dots we get $\langle m_{\rm N} \rangle = 0.57$, $\mu_{\rm N} = 1.20$.  The $\chi^{2}_{\nu}$ per degree of freedom is equal to 3.5 (black dots) and 3.2 (red dots). The gray line at $R = 1$ represents the boundary between classical and nonclassical correlations.} 
\label{RvsTWBTh}
\end{figure} 
The same considerations are also valid in the case of a thermal noise source: in Fig.~\ref{RvsTWBTh} we plot the noise reduction factor as a function of the mean number of photons of TWB for a fixed choice of the noise source. The data are shown as black dots + error bars in the absence of noise and as red dots + error bars in the presence of it. It can be easily observed that for small mean values of the TWB state the observation of nonclassicality is prevented because of the noise source, while it is still possible to reveal sub-shot-noise correlations for large mean values.\\
This behavior suggests that, in the presence of noise sources, the exploitation of well-populated TWB states in communication channels must be preferred since they are more robust to losses than low-populated ones. Obviously, the nonclassicality degradation also depends on the mean value of the noise. The larger the noise contribution, the more populated the quantum state should be in order to preserve its nonclassicality. In order to shed more light on this point, we consider the two noise sources (coherent and thermal states) used in the experimental part and also the case of a Fock-state noise source since they represent three benchmark situations (Poissonian, superPoissonian and subPoissonian noise).\\ 
According to Eq.~(\ref{noisyRcoh}), if the noise contribution is Poissonian, sub-shot-noise correlations are guaranteed if 
\begin{equation} \label{ineqCOH}
\frac{2\eta t \langle m \rangle}{(1+t) \langle m \rangle + \langle m_{\rm N} \rangle} - \frac{(1-t)^2 \langle m \rangle^2/\mu}{ (1+t) \langle m \rangle + \langle m_{\rm N} \rangle} > 0.
\end{equation}
Note that in this inequality the term $\langle m_{\rm N} \rangle$ referred to the noise contribution only appears in the equal denominators of the two terms. Thus it does not contribute to the establishment of a threshold for nonclassicality.
By focusing on the two numerators and solving the inequality in the variable $t$, we have
\begin{equation}\label{ineq}
{2\eta t \mu} - {(1-t)^2 \langle m \rangle}> 0.
\end{equation}
Thus, the nonclassicality condition holds for all values of the transmission efficiency between $t_1$ and 1, $t_1$ being the value of the smallest solution of the second-order equation, namely
\begin{equation} \label{SMsol}
t_1 = \left( 1+ \frac{\eta \mu}{\langle m \rangle} \right) - \sqrt{\frac{\eta \mu}{\langle m \rangle} \left( 2+ \frac{\eta \mu}{\langle m \rangle} \right)}.
\end{equation}
In practical situations, such as for the experimental condition described above, the quantity $\mu \eta / \langle m \rangle$ is of the order of 10-20, thus meaning that only for values of $t<0.05$ the inequality $R<1$ is not satisfied. \\
At variance with the case of Poissonian noise, for a thermal noise contribution the nonclassicality condition $R<1$ is explicitly dependent on the mean value $\langle m_{\rm N} \rangle$. Indeed, from Eq.~(\ref{noisyRth}) we have sub-shot-noise correlations if  
\begin{equation} \label{ineqTH}
\frac{2\eta t \langle m \rangle - \langle m_{\rm N} \rangle^2/\mu_{\rm N}}{(1+t) \langle m \rangle + \langle m_{\rm N} \rangle} 
- \frac{(1-t)^2 \langle m \rangle^2}{\mu \left[(1+t) \langle m \rangle + \langle m_{\rm N} \rangle \right]}>0.
\end{equation}
By solving the inequality for the variable $\langle m_{\rm N} \rangle$, that is for the mean number of photons of the noise source,
we have that $R<1$ if
\begin{equation}\label{mTH}
\langle m_{\rm N} \rangle < \sqrt{\mu_{\rm N} \left[ 2 \eta t \mu - (1-t)^2 \langle m \rangle \right] \frac{\langle m \rangle}{\mu}}.
\end{equation}
It is interesting to notice that, also in the case of thermal noise, the condition in Ineq.~(\ref{ineq}) must be fulfilled to ensure the positivity of the square-root argument. In this condition, the maximum value of $\langle m_{\rm N} \rangle$ satisfying the nonclassicality condition is limited by the mean number of photons per mode of the TWB state, $\langle m \rangle / \mu$. \\
Finally, we consider the case of a Fock-state noise source, which represents the best subPoissonian state.
According to Eq.~(\ref{noisyRfock}), the nonclassicality condition $R<1$ is satisfied for 
\begin{equation} \label{ineqFOCK}
\frac{\eta t (2 \langle m \rangle + \langle m_{\rm N} \rangle)}{(1+t) \langle m \rangle + \langle m_{\rm N} \rangle} - \frac{(1-t)^2 \langle m \rangle^2}{\mu \left[(1+t) \langle m \rangle + \langle m_{\rm N} \rangle \right]} > 0,
\end{equation}
which gives the condition
\begin{equation}\label{mFOCK}
\langle m_{\rm N} \rangle > - \left[ 2 - \frac{(1-t)^2 \langle m \rangle}{ \eta t \mu} \right] \langle m \rangle.
\end{equation}
Again, we see that for $t > t_1$ in Eq.~(\ref{SMsol}) there is no threshold for $\langle m_{\rm N} \rangle$, while for $t<t_1$ the threshold is set by Ineq.~(\ref{ineq}).\\
We can summarize the results obtained for the three situations as follows:
\begin{itemize}
\item In the case of coherent noise (Poissonian case), the noise contribution never prevents the observation of sub-shot-noise correlations but the value of the transmittance efficiency $t$ must exceed a threshold. 
\item In the case of thermal noise (superPoissonian case), the noise contribution can prevent the observation of sub-shot-noise correlations. In particular, the mean value of noise is upper bounded by the square root of the mean number of photons per mode of the TWB state. Moreover, the observation of nonclassicality also depends on the same threshold for the value of the transmittance efficiency $t$ as in the case of Poissonian noise. 
\item In the case of Fock-state noise (subPoissonian case), there is a condition for the mean value of the Fock state connected to the already-mentioned threshold for the value of the transmittance efficiency. If $t$ is larger than this threshold, the noise contribution never prevents the observation of sub-shot-noise correlations. On the contrary, for values of $t$ smaller than the threshold, only large mean values of the Fock state allow reaching the condition $R<1$.
\end{itemize}
All these results make the use of mesoscopic quantum states of light particularly appealing for Quantum-state Communication, as an alternative to the traditionally-employed single-photon states. 

\section{Conclusions}

In this paper we presented a comprehensive model to quantify the nonclassicality of mesoscopic twin beam states when one party is affected by noise and loss. In particular, we showed that the more super-Poissonian the noise contribution the more difficult the observation of nonclassicality. We also demonstrated that subPoissonian states minimally affect the nonclassicality level and a Fock state can preserve it. Our investigation was developed in terms of the noise reduction factor for measurable quantities in order to directly apply the theory to experimental data. To prove the model, we realized two experimental tests by using two different noise sources, namely a coherent state and a quasi-single-mode thermal state. The data are in excellent agreement with the model. Our results suggest the employment of mesoscopic twin-beam states in communication channels. In particular, well-populated TWB states seem to be the best choice in the presence of a strong noise contribution since they preserve their nonclassicality. Of course, the determination of photon numbers in rather intense states can be a problem. The best choice for the detectors could be represented by Silicon photomultipliers, which are compact photon-number-resolving detectors characterized by a high dynamic range \cite{cassina21}. Work is in progress in this direction. Finally we note that, since the developed model allow us to obtain information both about the quantum state and the noise sources, it could be used to describe a communication protocol based on TWB states in which the information is encoded in the noise superimposed to the TWB and the losses mimic an evesdropper's interference \cite{manu}.
\\
\acknowledgements A. A. acknowledges the Project “Investigating the effect of noise sources in the free-space
transmission of mesoscopic quantum states of light” supported by the University of Insubria.


\begin{thebibliography}{}

\bibitem{Mar03}  I. Marcikic, H. de Riedmatten, W. Tittel, H. Zbinden, and N. Gisin ``Long-distance teleportation of qubits at telecommunication wavelengths,'' Nature {\bf 421}, 509 (2003).  
\bibitem{Mul95} A. Muller, H. Zbinden and N. Gisin, ``Underwater quantum coding,'' Nature {\bf 378}, 449 (1995).
\bibitem{Urs04} R. Ursin, T. Jennewein, M. Aspelmeyer, R. Kaltenbaek, M. Lindenthal, P. Walther, and A. Zeilinger , ``Quantum teleportation across the Danube,'' Nature {\bf 430}, 849 (2004). 
\bibitem{Kur02} C. Kurtsiefer, P. Zarda, M. Halder, H. Weinfurter, P. M. Gorman, P. R. Tapster and J. G. Rarity, ``A step towards global key distribution,'' Nature {\bf 419}, 450 (2002). 
\bibitem{Asp03} M. Aspelmeyer, H. R. B$\ddot{\rm o}$hm, T. Gyatso, T. Jennewein, R. Kaltenbaek, M. Lindenthal, G. Molina-Terriza, A. Poppe, K. Resch, M. Taraba, R. Ursin, P. Walther, and A. Zeilinger, ``Long-Distance Free-Space Distribution of Quantum Entanglement,'' Science {\bf 301}(5633), 621-623 (2003). 
\bibitem{Jin10} X.-M. Jin, J.-G. Ren, B. Yang, Z.-H. Yi, F. Zhou, X.-F. Xu, S.-K. Wang, D. Yang, Y.-F. Hu, S. Jiang, T. Yang, H. Yin, K. Chen, C.-Z. Peng, and J.-W. Pan, ``Experimental free-space quantum teleportation,'' Nat. Photon. {\bf 4}, 376-381 (2010). 
\bibitem{Ji17} L. Ji, J. Gao, A.-L. Yang, Z. Feng, X.-F. Lin, Z.-G. Li, and X.-M. Jin, ``Towards quantum communications in free-space seawater,'' Opt. Express {\bf 25}(17), 19795-19806 (2017).
\bibitem{Huf20} F. Hufnagel, A. Sit, F. Bouchard, Y. Zhang, D. England, K. Heshami, B. J. Sussman, and E. Karimi, ``Investigation of underwater quantum channels in a 30 meter flume tank using structured photons,'' New J. Phys. {\bf 22}, 093074 (2020).
\bibitem{Vasy12} D. Yu. Vasylyev, A. A. Semenov, and W. Vogel, ``Toward global quantum communication: beam wandering preserves nonclassicality,'' Phys. Rev. Lett. {\bf 108}, 220501 (2012).
\bibitem{Capraro12}  I. Capraro, A. Tomaello, A. Dall’Arche, F. Gerlin, R. Ursin, G. Vallone, and P. Villoresi, ``Impact of turbulence in long range quantum and classical communications,'' Phys. Rev. Lett. {\bf 109}, 200502 (2012).
\bibitem{Boh17} M. Bohmann, R. Kruse, J. Sperling, C. Silberhorn, and W. Vogel, “Probing free-space quantum channels with laboratory-based experiments,” Phys. Rev. A {\bf 95}, 063801 (2017).
\bibitem{Cozzolino19} D. Cozzolino, B. Da Lio, D. Bacco, and L. K. Oxenl${\rm \o}$we, ``High-dimensional Quantum Communication: Benefits, progress, and future challenges,'' Adv. Quantum Technol. {\bf 2}, 1900038 (2019).
\bibitem{Flamini19} F. Flamini, N. Spagnolo, and F. Sciarrino, ``Photonic quantum information processing: a review,'' Rep. Prog. Phys. {\bf 82}, 016001 (2019).
\bibitem{But98}  W. Buttler, R. Hughes, P. Kwiat, S. Lamoreaux, G. Luther, G. Morgan, J. Nordholt, C. Peterson, and C. Simmons, ``Practical Free-Space Quantum Key Distribution over 1 km,'' Phys. Rev. Lett. {\bf 81}, 3283 (1998).
\bibitem{Mir15} M. Mirhosseini, O. S. Magaña-Loaiza, M. N. O’Sullivan, B. Rodenburg, M. Malik, M. P. Lavery, M. J. Padgett, D. J. Gauthier, and R. W. Boyd, ``High-dimensional quantum cryptography with twisted light,'' New J. Phys. {\bf 17} 033033 (2015).
\bibitem{Jin19} J. Jin, J.-P. Bourgoin, R. Tannous, S. Agne, C. J. Pugh, K. B. Kuntz, B. L. Higgins, and T. Jennewein, ``Genuine time-bin-encoded quantum key distribution over a turbulent depolarizing free-space channel,'' Opt. Express {\bf 27}(26), 37214-37223 (2019). 
\bibitem{JOSAB19} A. Allevi and M. Bondani, ``Preserving nonclassical correlations in strongly unbalanced conditions,'' J. Opt. Soc. Am. B {\bf 36}(12) 3275-3281 (2019).
\bibitem{ApplSci20} A. Allevi and M. Bondani, ``Tailoring asymmetric lossy channels to test the robustness of mesoscopic quantum states of light,'' Appl. Sci. {\bf 10}(24) 9094 (2020).
\bibitem{Weed10} C. Weedbrook, S. Pirandola, S. Lloyd, and T. C. Ralph, ``Quantum cryptography approaching the classical limit,'' Phys. Rev. Lett. {\bf 105}, 110501 (2010).
\bibitem{arimondo} A. Allevi and M. Bondani, ``Nonlinear and quantum optical properties and applications
of intense twin-beams,'' Adv. At. Mol. Opt. Phys. {\bf 66}, 49-110 (2017).
\bibitem{pra07} M. Bondani, A. Allevi, G. Zambra, M. G. A. Paris, and A. Andreoni, ``Sub-shot-noise photon-number correlation in a mesoscopic twin beam of light,'' Phys. Rev. A {\bf 76}(1), 013833 (2007).
\bibitem{jointdiff} A. Agliati, M. Bondani, A. Andreoni, G. De Cillis, and M. G. A. Paris, ``Quantum and classical correlations of intense beams of light investigated via joint photodetection,'' J. Opt. B Quantum Semiclassical Opt. {\bf 7}, S652-S663 (2005).
\bibitem{EPJD18} A. Allevi and M. Bondani, ``Can nonclassical correlations survive in the presence of asymmetric lossy channels?,'' Eur. Phys. J. D {\bf 72}, 178 (2018).
\bibitem{ASL} M. Bondani, A. Allevi, and A. Andreoni, ``Light Statistics by Non-Calibrated Linear Photodetectors,'' Adv. Sci. Lett. {\bf 2}, 463-468 (2009).
\bibitem{EPL} A. Allevi, A. Andreoni, F. A. Beduini, M. Bondani, M. G. Genoni, S. Olivares, and M. G. A. Paris, ``Conditional measurements on multimode pairwise entangled states from spontaneous parametric downconversion,'' Eur. Phy. Lett. {\bf 92}, 20007 (2010).
\bibitem{lau03} J. Laurat, T. Coudreau, N. Treps, A. Ma$\hat{\rm i}$tre, and C. Fabre, ``Conditional Preparation of a Quantum State in the Continuous Variable Regime: Generation of a sub-Poissonian State from Twin Beams,'' Phys. Rev. Lett. {\bf 91}(21), 213601 (2003).
\bibitem{our06} A. Ourjoumtsev, R. Tualle-Brouri, and P. Grangier, ``Quantum Homodyne Tomography of a Two-Photon Fock State,'' Phys. Rev. Lett. {\bf 96}(21), 213601 (2006).
\bibitem{perina13} J. Pe$\check{\rm r}$ina, O. Haderka, and V. Mich$\acute{\rm a}$lek, ``Sub-Poissonian-light generation by postselection from twin beams,'' Opt. Express {\bf 21}(16), 19387-19394 (2013). 
\bibitem{isk16} T. S. Iskhakov, V. C. Usenko, U. L. Andersen, R. Filip, M. V. Chekhova, and G. Leuchs, ``Heralded source of bright multi-mode mesoscopic sub-Poissonian light,'' Opt. Lett. {\bf 41}(10), 2149-2152 (2016).
\bibitem{lamperti} M. Lamperti, A. Allevi, M. Bondani, R. Machulka, V. Mich$\acute{\rm a}$lek, O. Haderka, and J. Pe$\check{\rm r}$ina Jr., ``Optimal sub-Poissonian light generation from twin beams by photon-number resolving detectors,'' J. Opt. Soc. Am. B {\bf 31}(1), 20-25 (2014).
\bibitem{OL19} G. Chesi, L. Malinverno, A. Allevi, R. Santoro, M. Caccia, and M. Bondani, ``Measuring nonclassicality with silicon photomultipliers,'' Opt. Lett. {\bf 44}(6), 1371-1374 (2019).
\bibitem{pra12} A. Allevi, S. Olivares, and M. Bondani, ``Measuring high-order photon-number correlations in experiments with multimode pulsed quantum states,'' Phys. Rev. A {\bf 85}(6), 063835 (2012).
\bibitem{cassina21} S. Cassina, A. Allevi, V. Mascagna, M. Prest, E. Vallazza, and M. Bondani, ``Exploiting the wide dynamic range of silicon photomultipliers for quantum optics applications,'' EPJ Quantum Technol. {\bf 8}, 4 (2021).
\bibitem{manu} A. Allevi and M. Bondani, ``Mesoscopic twin beams and photon-number-resolving detectors: A novel scheme for secure data transmission,'' manuscript in preparation.


\end{thebibliography}
\end{document}